\documentclass[twocolumn]{aastex631}
\usepackage{xspace}
\usepackage{wasysym}    

\newcommand\Msun{\text{M}_{\astrosun}} 
\newcommand\Zsun{\text{Z}_{\astrosun}} 
\newcommand\HI{\mbox{H\,\textsc{i}}\xspace} 
\newcommand\thesan{\mbox{\textsc{THESAN}}\xspace}
\newcommand\thesanhigh{\mbox{\textsc{THESAN-high}}\xspace}
\newcommand\thesanlow{\mbox{\textsc{THESAN-low}}\xspace}

\shorttitle{Bridging CD--EoR}
\shortauthors{Bera et al.}
\usepackage[caption=false]{subfig}

\begin{document}

\title{Bridging the Gap between Cosmic Dawn and Reionization favors Faint Galaxies-dominated Models}

\author[0000-0001-7072-570X]{Ankita Bera}
\altaffiliation{E-mail: \href{mailto: ankita1.rs@presiuniv.ac.in}{ankita1.rs@presiuniv.ac.in}}
\affiliation{School of Astrophysics, Presidency University, 86/1 College Street, Kolkata 700073, India}\affiliation{Center for Computational Astrophysics, Flatiron Institute, 162 5th Ave, New York, NY, 10010, USA}
\affiliation{Department of Astronomy, University of Maryland, College Park, MD 20742, USA}

\author[0000-0002-1050-7572]{Sultan Hassan}
\altaffiliation{NHFP Hubble Fellow.}
\affiliation{Center for Cosmology and Particle Physics, Department of Physics, New York University, 726 Broadway, New York, NY 10003, USA}
\affiliation{Center for Computational Astrophysics, Flatiron Institute, 162 5th Ave, New York, NY, 10010, USA}\affiliation{Department of Physics \& Astronomy, University of the Western Cape, Cape Town 7535,
South Africa}

\author[0000-0002-2838-9033]{Aaron Smith}
\affiliation{Center for Astrophysics $\vert$ Harvard $\&$ Smithsonian, 60 Garden Street, Cambridge, MA 02138, USA}
\affiliation{Department of Physics, The University of Texas at Dallas, Richardson, Texas 75080, USA}

\author[0000-0001-8531-9536]{Renyue Cen}
\affiliation{Institute for Advanced Study in Physics, Zhejiang University, Hangzhou 310027, China}
\affiliation{Institute of Astronomy, Zhejiang University, Hangzhou 310027, China}
\affiliation{Department of Astrophysical Sciences, Princeton University, Peyton Hall, Princeton, NJ, 08544, USA} 

\author[0000-0002-6021-7020]{Enrico Garaldi}
\affiliation{Max-Planck Institute for Astrophysics, Karl-Schwarzschild-Str.~1, D-85741 Garching, Germany}

\author[0000-0001-6092-2187]{Rahul Kannan}
\affiliation{Center for Astrophysics $\vert$ Harvard $\&$ Smithsonian, 60 Garden Street, Cambridge, MA 02138, USA} \affiliation{Department of Physics and Astronomy, York University, 4700 Keele Street, Toronto, Ontario MJ3 1P3, Canada}

\author[0000-0001-8593-7692]{Mark Vogelsberger}
\affiliation{Department of Physics, Massachusetts Institute of Technology, Cambridge, MA 02139, USA}





\begin{abstract}

It has been claimed that traditional models struggle to explain the tentative detection of the 21\,cm absorption trough centered at $z\sim17$ measured by the EDGES collaboration. On the other hand, it has been shown that the EDGES results are consistent with an extrapolation of a declining UV luminosity density, following a simple power-law of deep Hubble Space Telescope observations of $4 < z < 9$ galaxies.
We here explore the conditions by which the EDGES detection is consistent with current reionization and post-reionization observations, including the neutral hydrogen fraction at $z\sim6$--$8$, Thomson scattering optical depth, and ionizing emissivity at $z\sim5$. By coupling a physically motivated source model derived from radiative transfer hydrodynamic simulations of reionization to a Markov Chain Monte Carlo sampler, we find that it is entirely possible to reconcile the high-redshift (cosmic dawn) and low-redshift (reionization) existing constraints.
In particular, we find that high contribution from low-mass halos along with high photon escape fractions are required to simultaneously reproduce cosmic dawn and reionization constraints. Our analysis further confirms that low-mass galaxies produce a flatter emissivity evolution, which leads to an earlier onset of reionization with gradual and longer duration, resulting in a higher optical depth. While our faint-galaxies dominated models successfully reproduce the measured globally averaged quantities over the first one billion years, they underestimate the late redshift-instantaneous measurements in efficiently star-forming and massive systems. We show that our (simple) physically-motivated semi-analytical prescription produces consistent results with the (sophisticated) state-of-the-art \thesan radiation-magneto-hydrodynamic simulation of reionization.

\end{abstract}



\section{Introduction} \label{sec:intro}

Some of the most important but least explored phases in the evolutionary history of our Universe are the epochs of cosmic dawn and reionization. The ignition of the first luminous cosmic structures, either stars or active galactic nuclei (AGN), marks the beginning of the cosmic dawn. These sources of first light gradually and inhomogeneously transformed the intergalactic medium (IGM) from being predominately filled with cold neutral hydrogen gas to a warm ionized plasma \citep[for reviews see e.g.][]{2001PhR...349..125B, 2019arXiv190706653W, 2021arXiv211013160R}. 

Currently, there are several observational global constraints on reionization at $z\gtrsim5$, including measurements of the volume-averaged neutral hydrogen fraction of the IGM
using different techniques such as the evolution in Lyman-$\alpha$ and Lyman-$\beta$ forests \citep{2001AJ....122.2833F, White_2003, 2015MNRAS.447..499M} and Lyman-$\alpha$ emitters \citep{Ouchi_2010, 2011ApJ...726...38Z, 10.1093/mnras/stu2089, 10.1093/mnras/stv1751,2019MNRAS.485.3947M}, the optical depth to Thomson scattering of cosmic microwave background (CMB) photons as measured by \citet{2020A&A...641A...6P}, and the ionizing emissivity constraints as compiled by \citet{2013MNRAS.436.1023B}.  However, more stringent constraints are expected from the 21\,cm measurements, which will play a crucial role to map out the \HI distribution to trace the large-scale structure. Experiments such as the Low Frequency Array \citep[LOFAR,][]{2017ApJ...838...65P}, Hydrogen Epoch of Reionization Array \citep[HERA,][]{2017PASP..129d5001D}, and Square Kilometre Array \citep[SKA,][]{2013ExA....36..235M}, are devoted to detect the 21\,cm fluctuations during reionization.

Complementary constraints on cosmic dawn come from the measurements of the 21\,cm global signal at $z>15$. These include the Experiment to Detect the Global Reionization Signature \citep[EDGES,][]{2018Natur.555...67B}, Shaped Antenna measurement of the background Radio Spectrum \citep[SARAS,][]{Saurabh_2021}, Large-Aperture Experiment to Detect the Dark Ages \citep[LEDA,][]{2018MNRAS.478.4193P}, Radio Experiment for the Analysis of Cosmic Hydrogen \citep[REACH,][]{unknown, 2022NatAs...6..984D}. At this epoch, the controversial EDGES detection \citep{2018Natur.555...67B} of a flattened absorption profile in the sky-averaged radio spectrum, centered at 78\,MHz with an amplitude of 0.5\,K, provides a tentative constraint on the 21\,cm global signal during cosmic dawn.  Explaining the detection of such a deep absorption trough requires additional physics that can lead, for instance, to a higher SFR density \citep{Mirocha_2019, Mebane_2020, 2022MNRAS.515.2901M} or new exotic physics during cosmic dawn such as the interaction between dark-matter and baryons \citep[see e.g.,][]{Barkana18Nature, Munoz18, Slatyer2018}, and the excess radio background \citep[e.g.,][]{Fraser_2018, Pospelov_2018, Feng_2018, Ewall_2018, Fialkov_2019, Ewall2020}. 

Inferred from the EDGES signal, \citet{2018MNRAS.480L..43M} have derived a constraint on the early star formation and shown that it is consistent with an extrapolation of UV measurements at lower redshifts ($4 < z < 9$). While there have been many theoretical \citep[see e.g.,][]{Hills_2018, Bradley2019, Singh2019, Sims2020} and observational works \citep[see e.g.,][]{2022NatAs...6..607S} suggesting that the measured profile by EDGES is not of astrophysical origin, it is timely to understand what implications such a detection might have on cosmic dawn and reionization. In particular, could additional constraints be placed on the role of faint and bright galaxies during these epochs if the measurement holds? In light of the recent successful launch of the James Webb Space Telescope (JWST) and the growing observational efforts to detect the 21\,cm signal during cosmic dawn, many high-redshift constraints are expected, and hence it is currently a scientific priority to ask the question: {\it What is required to bridge the gap between cosmic dawn and reionization?} 

We here use a semi-analytical approach coupled with a Markov Chain Monte Carlo (MCMC) sampler to explore the range of scenarios and models that are mostly consistent with the combined constraints from cosmic dawn and reionization. Since all considered constraints are globally-averaged quantities, a semi-analytical approach is sufficiently accurate while also being much faster than using reionization simulations where numerical limitations in the resolution, box size, sub-grid physics, and radiation transport prescriptions complicate the interpretability and feasibility of parameter explorations \citep[e.g.][]{2011MNRAS.411..955M, 10.1093/mnras/stx420, 2020NatRP...2...42V, 2021JCAP...02..042W}. To obtain insights on the role of faint and bright galaxies, we use a physically motivated source model for the ionization rate ($R_{\rm ion}$) that was derived from radiative transfer simulations \citep{2015MNRAS.447.2526F, 2016MNRAS.457.1550H}. This parameterization assumes a non-linear relation between the ionization rate and the halo mass $R_{\rm ion} \propto M_h^{C+1}$, where $C=0$ corresponds to a linear relation, as is typically assumed in many semi-numerical models of reionization via the efficiency parameter \citep{2011MNRAS.411..955M}. In addition to the $C$ parameter, we aim to constrain the most debated parameter in reionization models, namely, the escape fraction of ionizing photons $f_{\rm esc}$. Harnessing the MCMC framework, our aim is to explore the joint $C$--$f_{\rm esc}$ parameter space to determine the range of models that naturally reproduce the combined constraints from reionization and cosmic dawn. Finally, we aim to study the implications these constraints might have on galaxy evolution during these early formation epochs.

This paper is organized as follows: we describe our empirical source model in \S\ref{sec:source} and calibrate it to cosmic dawn (EDGES) constraints in \S\ref{sec:mcd}. We then constrain our model to several reionization observables in \S\ref{sec:eor}, jointly to both cosmic dawn and reionization in \S\ref{sec:cdEoR}, and finally draw our concluding remarks in \S\ref{sec:conc}.

We assume a flat, $\Lambda$CDM cosmology throughout this paper with the cosmological parameters obtained from the recent \citet{2020A&A...641A...6P} measurements, i.e. $\Omega_{\Lambda} = 0.69$, $\Omega_{\rm m} = 0.31$, $\Omega_{\rm b} = 0.049$, and the Hubble parameter $H_0 = 67.66$~km/s/Mpc.

\begin{figure*}
    \centerline{\includegraphics[width=0.45\textwidth]{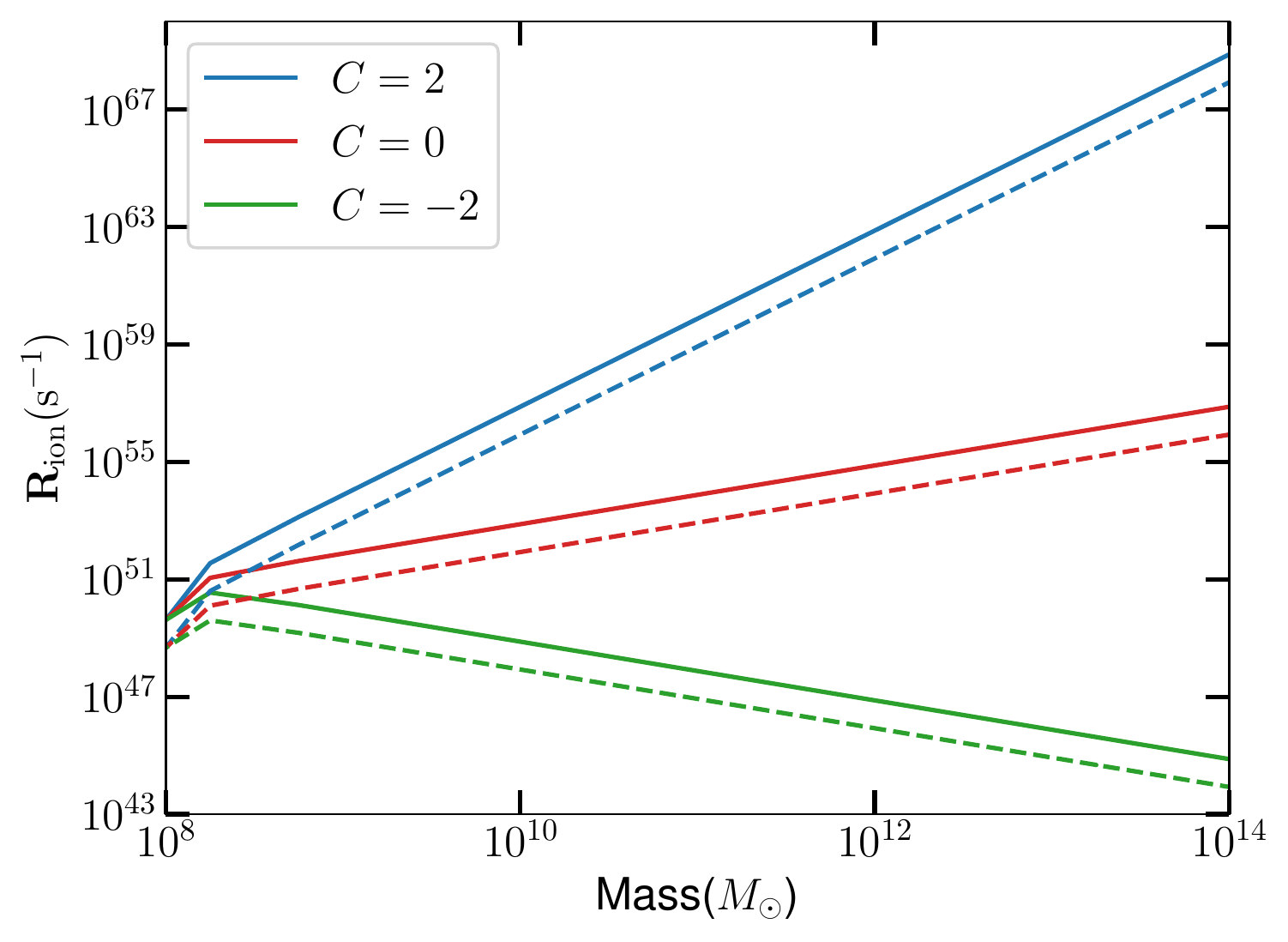}\includegraphics[width=0.45\textwidth]{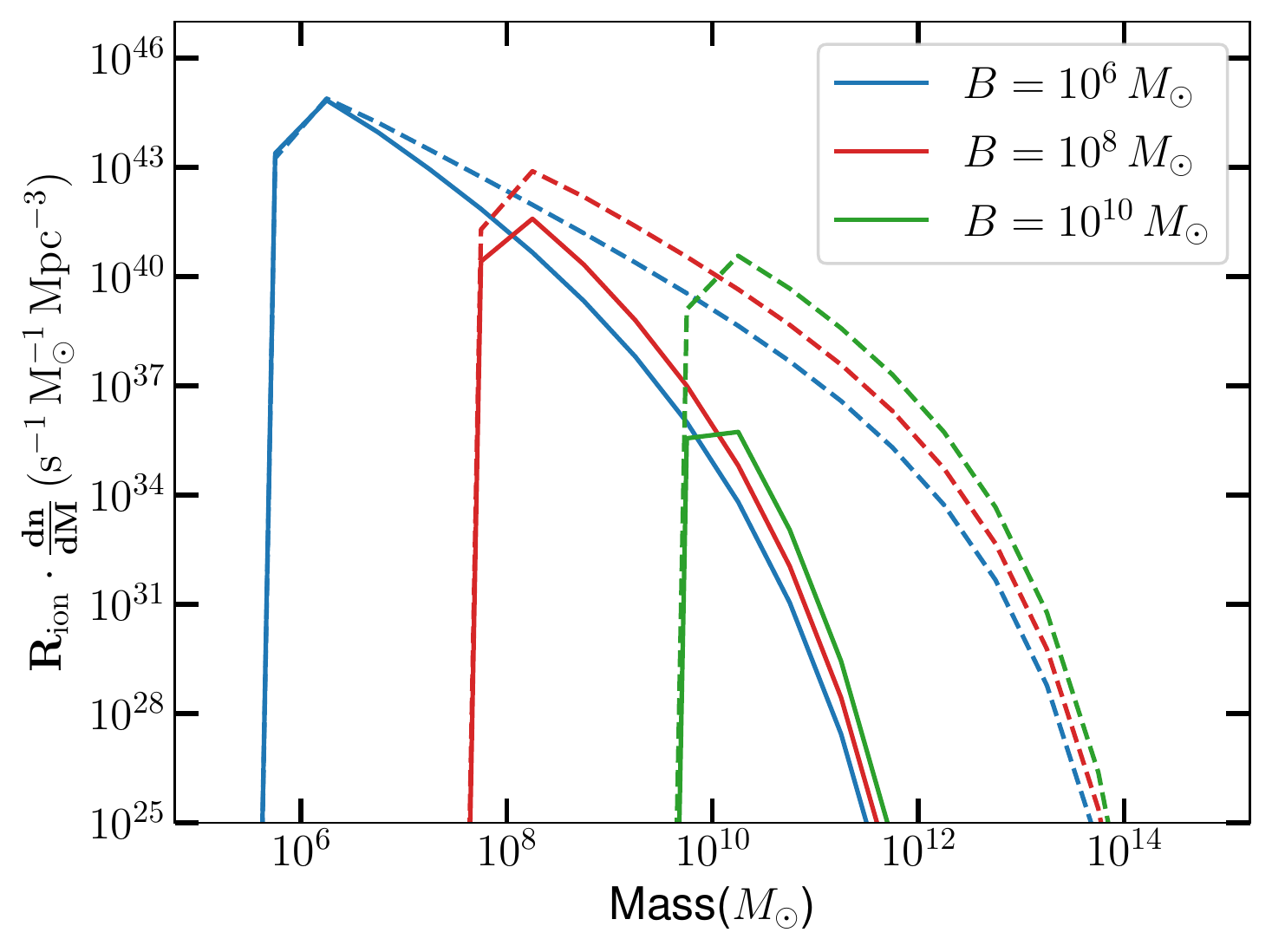}}
    \caption{{\bf Left panel:} The ionization rates for different $C$ values at $z\sim17$ (solid) and $z\sim6$ (dashed) while fixing other parameters to values of $A=10^{40}\,\Msun^{-1}\text{s}^{-1}$, $B=10^{8}\,\Msun$, and $D=2.28$ as derived from radiative transfer simulations \citep{2016MNRAS.457.1550H}. This shows that positive and negative $C$ correspond to higher ionization rate efficiencies from massive and low-mass halos, respectively. {\bf Right panel:} The ionization rates multiplied by the Sheth--Tormen halo mass function for different values of $B$, while keeping the other parameters fixed at $A=10^{40}\,\Msun^{-1}\text{s}^{-1}$, $C=-0.5$, and $D=2.28$. The $B$ parameter represents the minimum halo mass ($M_{h,\rm min}$) considered in the model.}
    \label{fig:Rion}
\end{figure*}

\section{Source model}
\label{sec:source}
In this section, we discuss the details of the source model considered in this work. To study the role of different populations of galaxies, we consider a physically motivated source model that was derived in \citet{2016MNRAS.457.1550H} from the radiative transfer simulations of reionization described in \citet{2018MNRAS.480.2628F}. In this model, the ionization rate $R_{\rm ion}$ is parameterized as a function of halo mass ($M_h$) and redshift ($z$). The mass dependence of $R_{\rm ion}$ is analogous to the Schechter function, which can be characterized as a power law on the bright end and an exponential cut-off on the faint end. The redshift dependence follows a simple power law.
This parameterization of $R_{\rm ion}$ accounts for the non-linear dependence on halo mass, which can be expressed as follows:
\begin{equation} \label{eq:Rion}
    \frac{R_{\rm ion}}{M_h} = A\,(1+z)^D \left(\frac{M_h}{B}\right)^C \exp\left[-\left(\frac{M_h}{B}\right)^{-3}\right] \, ,
\end{equation}
where $A$, $B$, $C$, and $D$ are free parameters, and their best-fit values were obtained by calibrating to radiative transfer simulations. We refer the reader to \citet{2016MNRAS.457.1550H} for more details about the derivation of this model. We now discuss the physical meaning of these free parameters. Parameter $A$ acts as an amplitude of $R_{\rm ion}$, which scales the ionizing emissivity over the entire halo mass range at a given redshift by the same amount. {Parameter $B$ determines the minimum halo mass, which can be thought of as the quenching mass scale due to feedback from star formation and photoionization heating.} Parameter $C$ quantifies the slope of the $R_{\rm ion}$--$M_h$ relation, which controls the contribution of different mass scales to the total emissivity. Lastly, parameter $D$ accounts for the redshift dependence of ionization rate for a given halo mass.

Having defined our source model $R_{\rm ion}$, we adopt the \citet{1999MNRAS.308..119S} halo mass function ($\frac{\text{d}n}{\text{d}M}$), which provides the number density of halos per unit halo mass, to compute the cosmic evolution of the global quantities governing reionization and comic dawn. In this work, we consider the halo mass range from $10^{5}$ to $10^{15}\,\Msun$ to account for the contribution from all source populations including the faintest halos.

The most interesting parameters of this source model are the $B$ and $C$ parameters, since they allow us to draw conclusions about the role of different source populations during cosmic dawn and reionization. 
Therefore, we set the amplitude $A = 10^{40}\,\Msun^{-1}\text{s}^{-1}$ and redshift index $D = 2.28$, following the calibration to radiative transfer simulations \citep[for more details see][]{2016MNRAS.457.1550H}\footnote{We have varied the $A$ and $D$ parameters to reproduce several observations using MCMC and found similar values. Hence, fixing these parameters has a minimal impact on the presented results.} throughout. To illustrate how the source model parameters ($B$ and $C$) affect the $R_{\rm ion}$--$M_h$ relation, in Figure~\ref{fig:Rion} we show several $R_{\rm ion}$ models with different $C$ and $B$ values at cosmic dawn ($z\sim17$, solid) and reionization ($z\sim6$, dashed). 
As the $C$ parameter scales the ionization rate ($R_{\rm ion}$) as a function of halo mass following $R_{\rm ion} \propto M_h^{C+1}$, $C=0$ (red curves) corresponds to a linear relation, shown in the left panel of Figure~\ref{fig:Rion}.
However, in the case of $C = 2$ (blue curves), the emissivity increases with halo mass and hence favors a relatively larger contribution from more massive halos than low-mass halos, and vice versa for the case of $C = -2$ (green curves), where reionization is dominated by low-mass halos. {While this depends on the photon escape fraction, under the assumption of a mass-independent escape fraction, 
the intrinsic $R_{\rm ion}$ increases by several orders of magnitude from low to high mass halos as seen in Figure~\ref{fig:Rion}.} In the right panel, we show the impact of varying the $B$ parameter, which sets the minimum halo mass scale. 

\begin{figure*}
    \centerline{\includegraphics[width=0.45\textwidth]{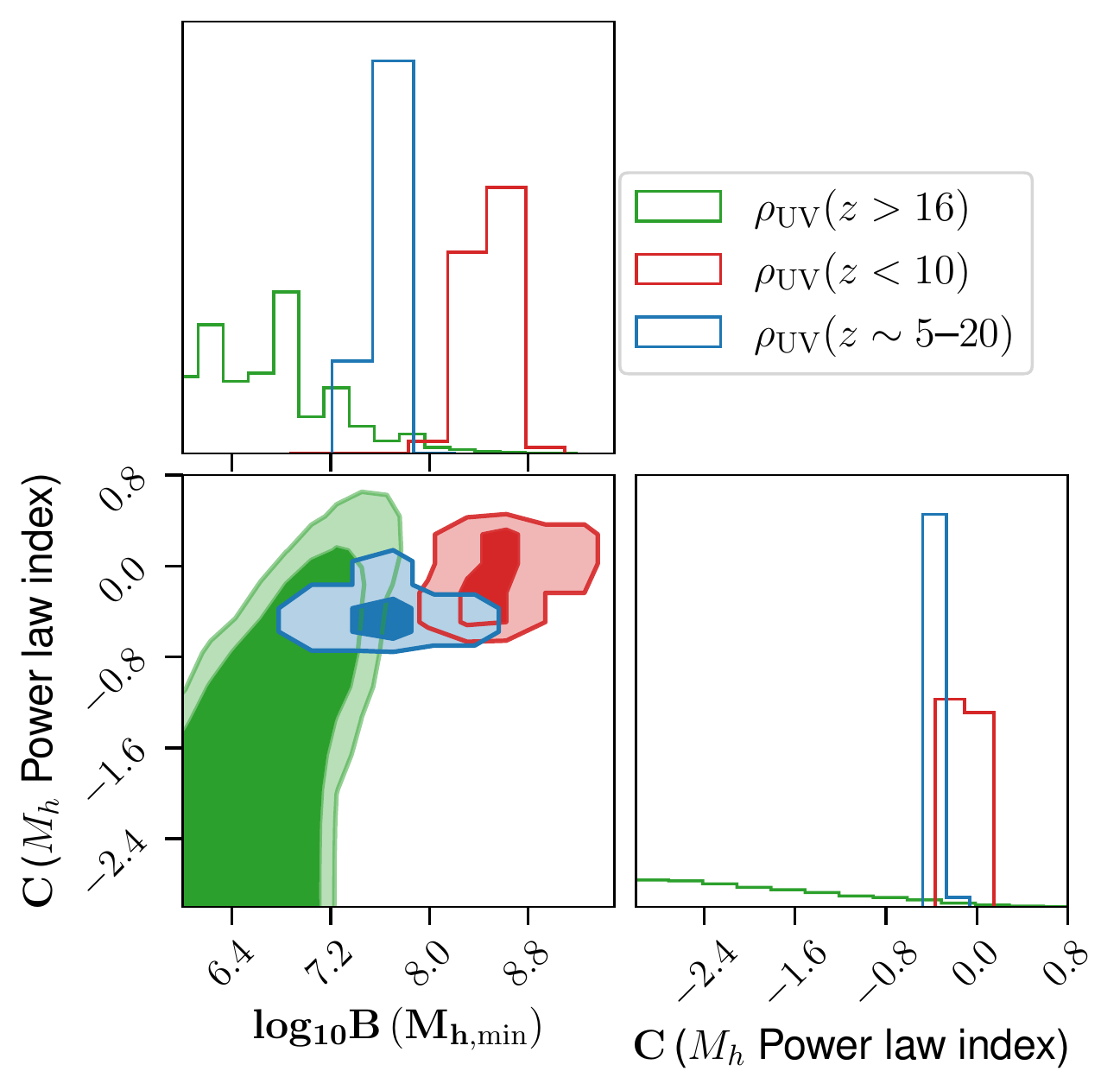}\includegraphics[width=0.45\textwidth]{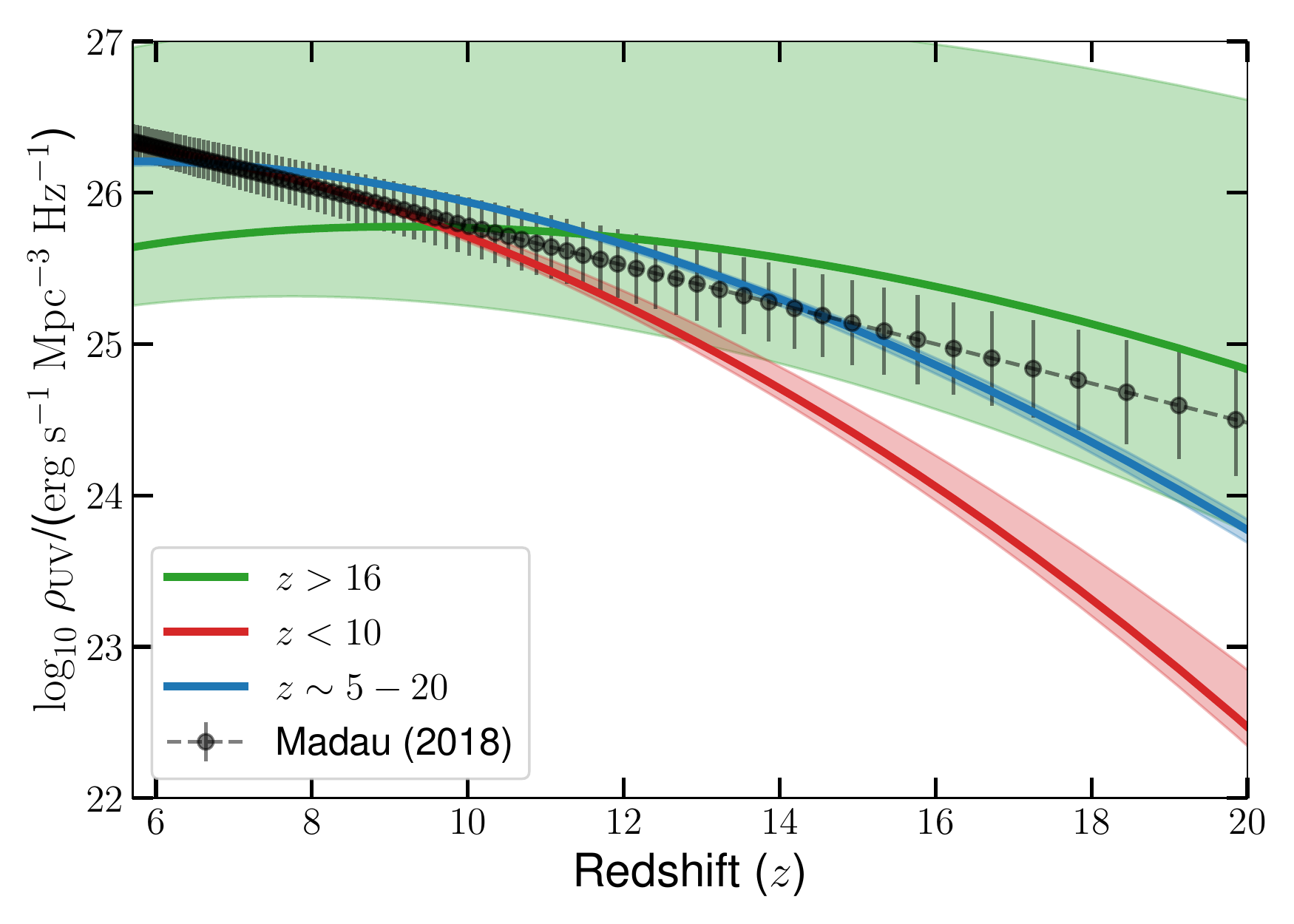}}
    \caption{{\bf Left panel:} Several posterior distributions derived using the UV luminosity density \citep{2018MNRAS.480L..43M} including constraints for three different redshift ranges, namely $z>16$ (representing cosmic dawn only, green), $z<10$ (during reionization only, red), and $z\sim5$--$20$ (cosmic dawn and reionization, by combining the full redshift range, blue). {\bf Right panel:} Comparison between several UV luminosity densities calculated from $R_{\rm ion}$ with the inferred parameter values using MCMC. This figure indicates that fitting to $z>16$ favors more negative $C$ values, which implies stronger contributions from low-mass (faint) galaxies.}
    \label{fig:SFR}
\end{figure*}

\section{Source model calibration to cosmic dawn (EDGES) constraints}
\label{sec:mcd}
%
Our goal is to find the possible range of models that can reproduce the existing observational constraints during cosmic dawn and reionization. To do so, we first calibrate our source model parameters ($C, B$) to the intrinsic UV luminosity density ($\rho_{\rm UV}$) inferred from the EDGES signal in the redshift range $z \sim 5$--$20$, where the function $\log_{10} [\rho_{\rm UV}/({\rm erg\,s^{-1}\,Mpc^{-3}\,Hz^{-1}})] = (26.30 \pm 0.12) + (-0.130 \pm 0.018)(z-6)$ is the best fit as compiled by \citet{2018MNRAS.480L..43M}.

Using our source model, we compute the UV luminosity density as follows. First, the ionization rate $R_{\rm ion}$ (Equation~\ref{eq:Rion}) can be converted to a star-formation rate (SFR) by accounting for a stellar metallicity-dependent parameterization of the ionizing photon flux $Q_{\rm ion}$ following:

\begin{equation} \label{eq:SFR}
    {\rm SFR} = R_{\rm ion}(M_h,z)/ Q_{\rm ion}(Z) \, .
\end{equation}
The sub-solar metallicity-dependent parameterization is provided by \citet{2011ApJ...743..169F}:
\begin{equation}
    \log Q_{\rm ion}(Z) = 0.639 (-\log Z)^{1/3} + 52.62 - 0.182 \, ,
\end{equation}
where $Q_{\rm ion}$ is in units of $\rm s^{-1} \, (\Msun\,yr^{-1})^{-1}$. This form of $Q_{\rm ion}$ is consistent with the equilibrium values measured by \citet{2003A&A...397..527S} assuming a \citet{2003PASP..115..763C} initial mass function (IMF). We also adopt the \citet{2017ApJ...840...39M} redshift evolution of the mass-weighted metallicity ($Z$) given by:
\begin{equation}
    \log\left(Z/\Zsun\right) = 0.153 - 0.074\,z^{1.34} \, .
\end{equation}
The SFR can then be written in terms of the UV luminosity density $L_{\rm UV}$ \citep{1998ARA&A..36..189K, 1998ApJ...498..106M} as follows:
\begin{equation} \label{eq:rho_UV}
    {\rm SFR} [{\rm \Msun\,yr^{-1}}] = 1.25 \times 10^{-28} L_{\rm UV} [{\rm ergs\,s^{-1}\,Hz^{-1}}] \, .
\label{eq:UVtoSFR}
\end{equation}
The global $\rho_{\rm UV}$ is obtained by integrating $\int L_{\rm UV}\frac{\text{d}n}{\text{d}m}\text{d}m$ over the entire mass range at different redshifts.

We now constrain our source model parameters ($B$ and $C$) to the \citet{2018MNRAS.480L..43M} UV luminosity evolution using {\sc emcee}, which is an affine-invariant ensemble sampler for MCMC \citep{2013PASP..125..306F}. We assume a flat prior in the range: $\log (B/\Msun) \in [5,10]$ and  $C \in [-5,5]$.
Our task is to find the range of models that minimizes the following multivariate $\chi^{2}$ distribution:
\begin{equation}
    \chi^{2} {\equiv} \sum_{z} \frac{(\rho_{\rm UV, model}(z) - \rho_{\rm UV,obs}(z))^{2}}{2\,(\sigma_{\rho_{\rm UV, obs}}(z))^{2}} \, .
    \label{eq:chi2}
\end{equation}

\begin{deluxetable}{cccc}
\tablenum{1}
\tablecolumns{4}
\tablecaption{Best-fit values of mass cut-off ($B$) and power dependence on halo mass ($C$).}
\tablewidth{0pt}
\tablehead{
\colhead{Parameters} & \colhead{$z>16$} & \colhead{$z<10$} & \colhead{$z\sim5-20$}
}
\startdata
\vspace{0.2cm} \\
${\bf \log_{10}B}$ & $6.34^{+0.69}_{-0.80}$ & $8.49^{+0.16}_{-0.17}$ & $7.67^{+0.07}_{-0.15}$ \\ \\ \hline
\vspace{0.2cm} \\
${\bf C}$ & $-6.43^{+3.78}_{-9.0}$ & $-0.11^{+0.04}_{-0.05}$ & $-0.34^{+0.03}_{-0.03}$ \\
\vspace{0.2cm}
\enddata
\tablecomments{The other model parameters are fixed to values obtained by fitting to radiative transfer simulations \citep[$A=10^{40}\,\Msun^{-1}\text{s}^{-1}$ and $D=2.28$, see][]{2016MNRAS.457.1550H}.}
\label{table:BandC}
\end{deluxetable}

In the left panel of Figure~\ref{fig:SFR} we show several posterior distributions of the parameters $B$ and $C$ by considering different epochs, namely, cosmic dawn ($z>16$, green), reionization ($z<10$, red), and the combination of cosmic dawn plus reionization ($z\sim 5$--$20$, blue). The dark and light shaded contours correspond to 1$\sigma$ and 2$\sigma$ levels, respectively. From the 1-dimensional probability distribution function (PDF) of the $B$ parameter (i.e. $M_{h,\rm min}$) it is clear that the minimum halo mass or the mass cut-off varies between $\sim 10^{6-8}\,\Msun$. It is also evident that calibrating the model to reproduce higher-$z$ $\rho_{\rm UV}$ constraints ($z > 16$) favors models with lower mass cut-offs ($\sim 10^{6}\,\Msun$) and vice versa. Likewise, a more negative $C$ is preferred to match with higher-$z$ constraints. As discussed before, the negative value of $C$ and the low value of $B$ both suggest that the low-mass (faint) halos play an important role in reproducing the extrapolated $\rho_{\rm UV}$ constraints. Overall, using $\rho_{\rm UV}$ measurements in the entire redshift range ($z\sim5$--$20$) provides much tighter posterior contours than constraining to either $z>16$ or $z<10$. This is mainly due to the number of redshift bins used in each case. Since the time duration in the redshift range of $z=16$--$20$ is shorter than that of $z=5$--$20$ or $z=5$--$10$, more redshift bins exist in the latter than the former, and hence more constraining power is expected. Each of the best-fit values with 1$\sigma$ errors is provided in Table~\ref{table:BandC} for the three scenarios. In the right panel of Figure~\ref{fig:SFR}, we compare the $\rho_{\rm UV}$ evolution predictions from these three models with the \citet{2018MNRAS.480L..43M} $\rho_{\rm UV}$ evolution. Shaded areas show the 1$\sigma$ confidence levels which are obtained by translating the parameter 1$\sigma$ levels from Table~\ref{table:BandC} into constraints on $\rho_{\rm UV}$. All models are within the 1$\sigma$ level of the $\rho_{\rm UV}$ constraints over the relevant redshift ranges. This figure clearly demonstrates that a stronger contribution from low-mass halos is required to reproduce the $\rho_{\rm UV}$ constraints inferred from EDGES during cosmic dawn (green curve). 
This can be seen in the right panel of Figure~\ref{fig:Rion}, where the maximum halo mass during reionization ($z=6$, dashed) is about 2 orders of magnitude higher than during cosmic dawn ($z=17$, solid).

Since our aim is to bridge the gap between cosmic dawn and reionization, we will fix the $B$ (i.e. $M_{h,\rm min}$) parameter to the value obtained by calibrating the source model to the $\rho_{\rm UV}$ constraints over the entire redshift range ($z=5$--$20$). At the same time, we will keep the halo mass power-law index parameter ($C$) as a free parameter to explore its correlations with the escape fraction of ionizing photons ($f_{\rm esc}$), and to test whether similar 
values can be derived by adding reionization constraints, which we discuss next.

\begin{figure*}
    \centering
    \includegraphics[width=0.33\textwidth]{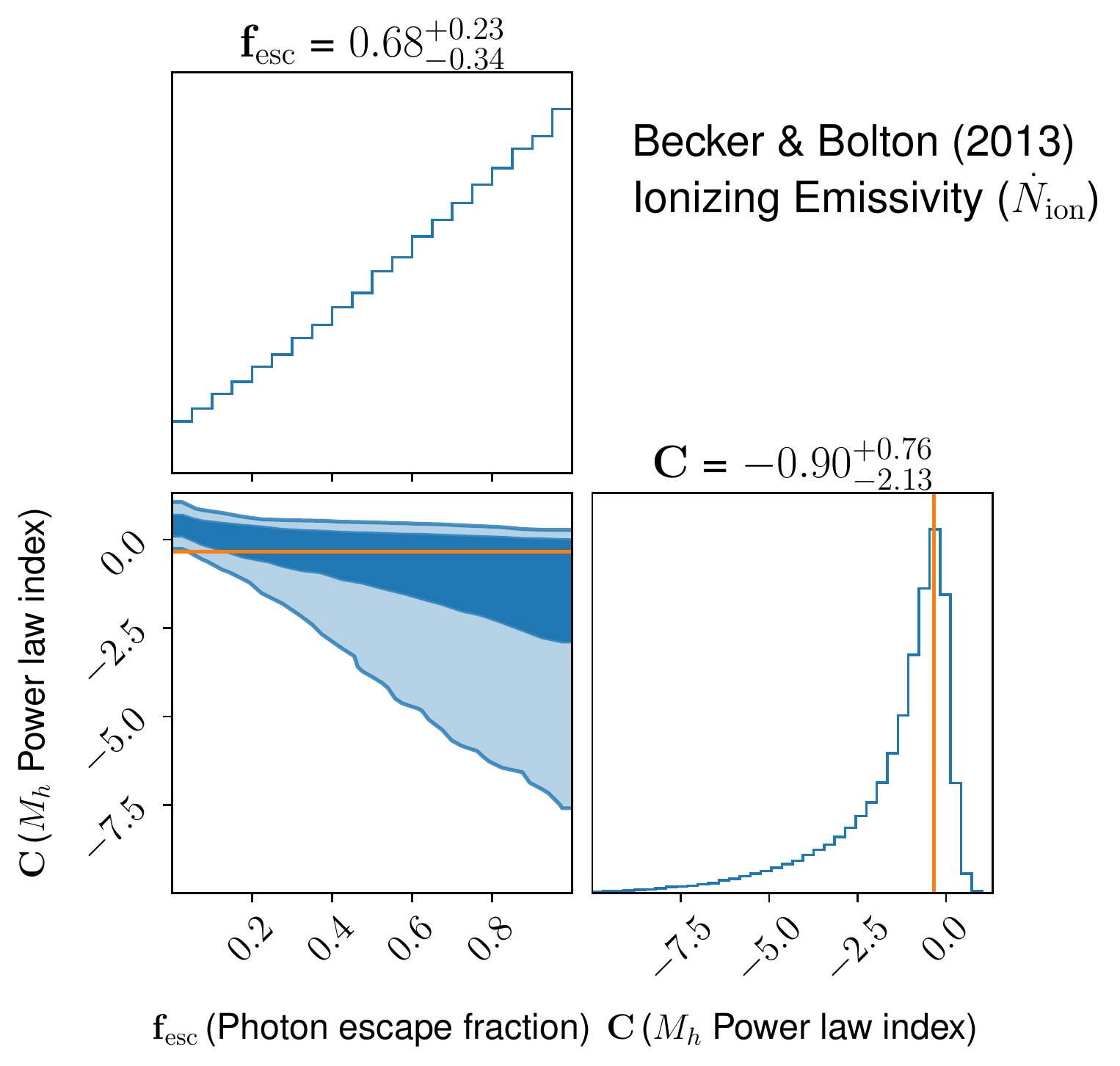}
    \includegraphics[width=0.3\textwidth]{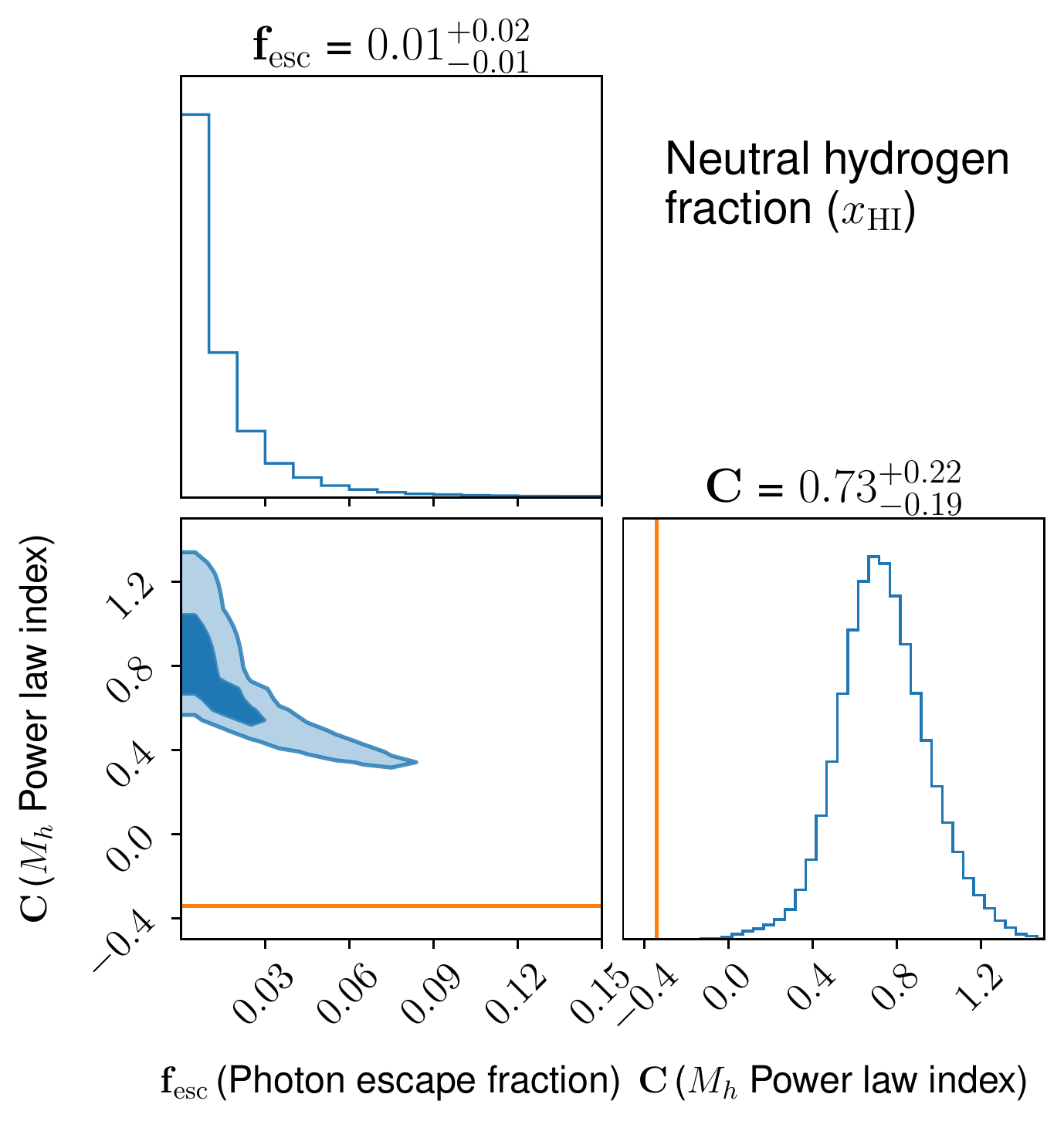}
    \includegraphics[width=0.3\textwidth]{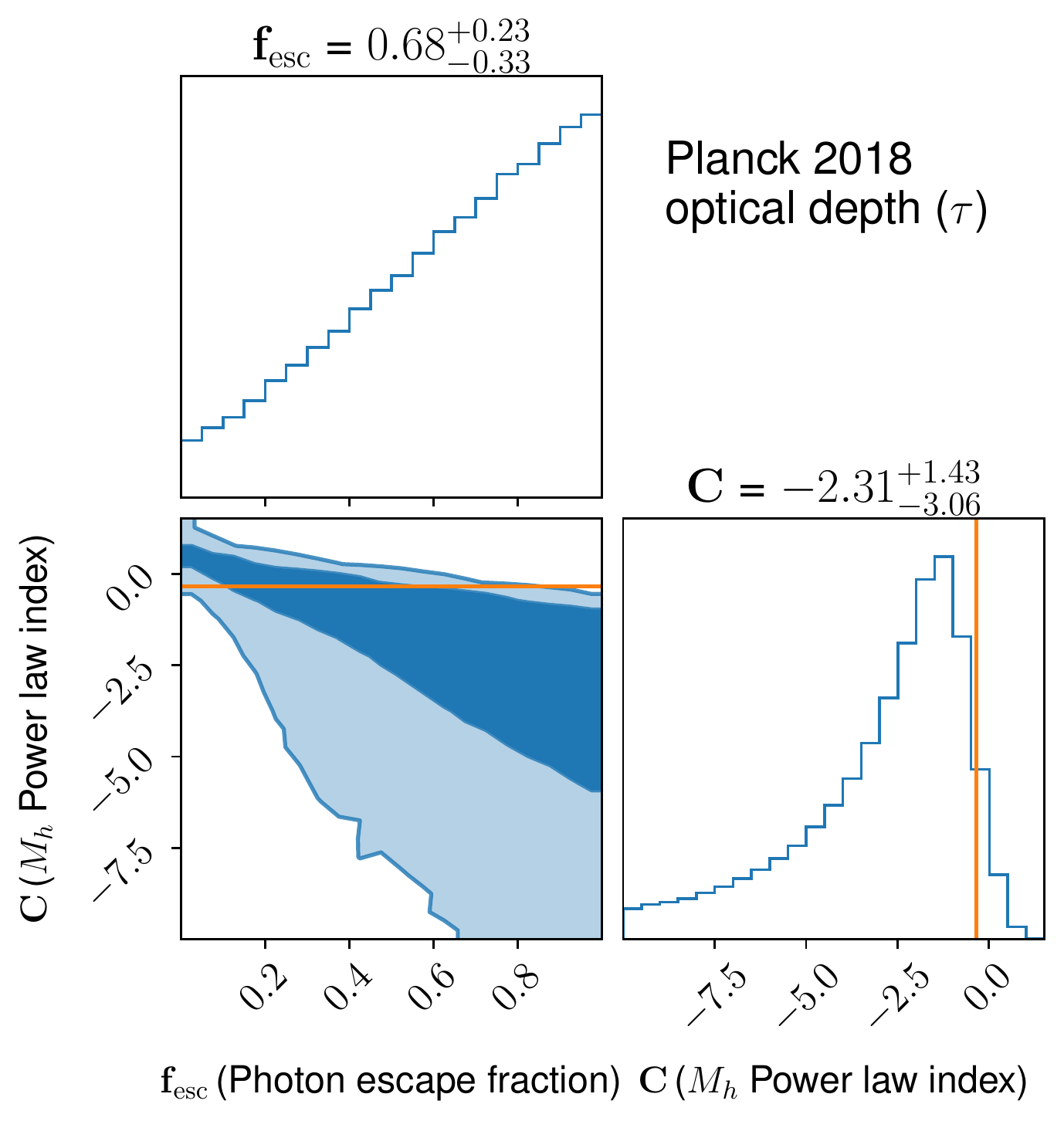}
    \caption{Posterior distributions for the escape fraction of ionizing photons $f_{\rm esc}$ and power-law index $C$ as constrained by the ionizing emissivity ($\dot{N}_{\rm ion}$; left), IGM neutral fraction ($x_{\rm HI}$; middle), and Thomson optical depth ($\tau$; right). {The orange line in each plot denotes the value of $C$ parameter obtained in Table~\ref{table:BandC} for $z \sim 5$--$20$, for reference.} The emissivity and optical depth do not provide tight constraints on our parameters, due to the large uncertainties. The IGM neutral fraction measurements provide tighter constraints on our parameters and favor models with low $f_{\rm esc}$ and positive $C$ values, suggesting that reionization is driven by massive (bright) galaxies.}
    \label{fig:eor_con}
\end{figure*}

\section{Constraining the source model for key reionization observables}
\label{sec:eor}
We consider three key reionization observables to place constraints on our source model parameters,
namely the ionizing emissivity, IGM neutral fraction, and Thomson optical depth to the CMB.
In addition to the halo mass power-law index parameter ($C$) we also vary
the fraction of ionizing photons ($f_{\rm esc}$) that successfully escapes the remaining neutral hydrogen clumps and dust extinction in the interstellar medium (ISM) and circumgalactic medium (CGM) to contribute to the IGM ionization process. In this work, we assume a constant $f_{\rm esc}$ at all redshifts, since our $R_{\rm ion}$ function (Equation~\ref{eq:Rion}) already accounts for redshift and mass dependence.
We assume a flat prior in the range: $f_{\rm esc} \in [0,1]$ and  $C \in [-10,10]$, and combine different data using a multivariate Gaussian likelihood (similar to Equation~\ref{eq:chi2}).

\subsection{The ionizing emissivity, \texorpdfstring{$\dot{N}_{\rm ion}$}{Ndot}}
The integrated emission rate density of ionizing photons, or ionizing emissivity ($\dot N_{\rm ion}$), is a measure of the total number of ionizing photons per second per volume that escape from all ionizing sources to the IGM. In our case, the ionizing emissivity is related to the ionization rate ($R_{\rm ion}$) as follows:
\begin{equation} \label{eq:Nion}
    \dot{N}_{\rm ion} [{\rm s^{-1} Mpc^{-3}}] = f_{\rm esc}  \int R_{\rm ion}(M_h, z) \frac{\text{d}n}{\text{d}M_h} \text{d}M_h \, ,
\end{equation}
where $\frac{\text{d}n}{\text{d}M_h}$ is the \citet{1999MNRAS.308..119S} differential halo mass function, which gives the number density of halos in the mass range of $M$ and $M+\text{d}M$ per unit comoving volume.
We constrain our source model parameters ($C, f_{\rm esc}$) to the \citet{2013MNRAS.436.1023B} ionizing emissivity constraint at $z=4.75$ of $\log_{10} [\dot{N}_{\rm ion}/({\rm 10^{51}\,photons\,s^{-1}\,Mpc^{-3}})] = -0.014^{+0.454}_{-0.355}$.

In the left panel of Figure~\ref{fig:eor_con}, we show the $f_{\rm esc}$--$C$ joint posterior distribution constrained to match the ionizing emissivity measurement.
This shows that the $\dot{N}_{\rm ion}$ measurement alone cannot place a tight constraint on these parameters due to the large uncertainty. However, $\dot{N}_{\rm ion}$ data favors models with higher $f_{\rm esc}$ and more negative $C$, leading to a stronger contribution from low-mass halos. Similar parameter constraints were found in our earlier work on calibrating semi-numerical simulations to reionization observables \citep{2017MNRAS.468..122H}. {The best-fit value of $C$ parameter obtained in the previous section for $z \sim 5$--$20$ is also within the $1\sigma$ level as shown by orange horizontal line in Figure~\ref{fig:eor_con}.}

\begin{deluxetable*}{cccc}
\tablenum{2}
\label{table:xHI_obs}
\tablecolumns{4}
\tablecaption{IGM neutral hydrogen fraction measurements}
\tablewidth{0pt}
\tablehead{
\colhead{Redshift($z$)} & \colhead{Constraints} & \colhead{Observables} & \colhead{References}
}
\startdata
\vspace{0.2cm} \\
${5.9}$ & $\leq 0.06\pm 0.05$ & Ly-$\alpha$ and Ly-$\beta$ forest dark fraction & \citet{2015MNRAS.447..499M} \\ \\ \hline
\vspace{0.2cm}  \\
${7.0}$ & $0.59^{+0.11}_{-0.15}$ & Ly-${\alpha}$ EW distribution & \citet{Mason_2018} \\ \\ \hline
\vspace{0.2cm} \\
${7.09}$ & $0.48 \pm 0.26$ &  & \citet{2018ApJ...864..142D} \\ \\
\vspace{0.2cm} 
${7.5}$ & $0.21^{+0.17}_{-0.19}$ & QSO damping wings & \citet{2019MNRAS.484.5094G} \\ \\ \vspace{0.2cm} 
${7.54}$ & $0.60^{+0.20}_{-0.23}$ &  & \citet{2018ApJ...864..142D} \\ \\ \hline
\vspace{0.2cm} \\
${7.6}$ & $0.88^{+0.05}_{-0.10}$ & Lyman-break galaxies emitting Ly-$\alpha$ & \citet{2019ApJ...878...12H} \\
\vspace{0.2cm}
\enddata
\end{deluxetable*}

\subsection{The IGM neutral fraction, \texorpdfstring{$x_{\rm HI}$}{xHI}}
\label{sec:xHI}
We compute the reionization history from our models as follows. The rate of change in the ionized fraction of intergalactic hydrogen ($x_{\rm HII}$) is given by \citep{1999ApJ...514..648M},
\begin{equation}
    \frac{\text{d}x_{\rm HII}}{\text{d}t} = \frac{\dot N_{\rm ion}}{\bar n_{\rm H}} - \frac{x_{\rm HII}}{\bar t_{\rm rec}} \, .
\end{equation}
The first term describes the growth as a ratio between comoving ionizing emissivity ($\dot N_{\rm ion}$) and volume-averaged comoving number density of intergalactic hydrogen $\bar n_{\rm H}$, which is given by
\begin{equation}
    \bar n_{\rm H} = X \Omega_{b,0} \rho_{{\rm crit},0} / m_{\rm H} \, .
\end{equation} 
Here, $X$ is the cosmic hydrogen mass fraction (0.76), $\rho_{{\rm crit},0}$ is the present-day critical density, and $m_{\rm H}$ is the mass of a hydrogen atom. The second term models the sink of ionizing photons, where the recombination time scale for the IGM is given by
\begin{equation}
    t_{\rm rec} = \left[ C_{\rm HII} \alpha_{\rm A} (1+\chi) {\bar n_{\rm H}} (1+z)^3 \right]^{-1} \, .
\end{equation}
Here, $C_{\rm HII}$ is the redshift-dependent clumping factor, which we adopt from \citet{2015MNRAS.451.1586P}. This clumping factor accounts for the overall density fluctuations in ionized medium, which boosts recombination rate by a factor up to $\sim 5$ near the end of reionization, predominantly contributed by ionized medium in the vicinity of halos. 
$\chi = Y/4X$, where, $Y$ denotes the helium mass fraction (0.24). Here, $\alpha_{\rm A}$ is the case~A recombination coefficient ($4.2 \times 10^{-13}\,{\rm cm^3\,s^{-1}}$), corresponding to a temperature of $10^4\,{\rm K}$ \citep{Kaurov_2014}.
Having computed the reionization history, we now constrain our source model parameters to the IGM neutral fraction measurements given in Table~\ref{table:xHI_obs}.

In the middle panel of Figure~\ref{fig:eor_con}, we show the parameter constraints given the above combination of IGM neutral fraction data. The $x_{\rm HI}$ data provides the tightest constraints,
with a clear tendency for the data to favor models with low $f_{\rm esc}$ and positive $C$ parameters. This implies that massive (bright) galaxies-dominated models are preferred. The low $f_{\rm esc}$ value (1\%) found here is a consequence of the non-linear relation between $R_{\rm ion}$ and $M_h^C$ through the $C$ parameter, where an anti-correlation between $f_{\rm esc}$ and $C$ is observed.
{However, we find that the resulting $C$ value is not within the range reported in Table~\ref{table:BandC}, hence we combine different constraints in the next section to understand the degeneracy between $f_{\rm esc}$ and $C$.}
Using the same source model in semi-numerical simulations of reionization, we have previously found that $f_{\rm esc}=4\%$ is sufficient to match the reionization observations \citep{2016MNRAS.457.1550H}, which is still within the 2$\sigma$ level of the current constraints based on updated observational data. We also find a lower $f_{\rm esc}$ value since our current analysis includes contributions from the wider halo mass range of $10^{5-15}\,\Msun$ as opposed to $10^{8-12}\,\Msun$ in \citet{2016MNRAS.457.1550H}. Such low $f_{\rm esc}$ values have been favored in several works \citep{gnedin2008escape, 2014MNRAS.442.2560W, 10.1093/mnras/stv1679,2019PhRvD..99b3518C,Rosdahl2022,Yeh2022}. 
However, it is worth noting that in this work, we consider a constant $f_{\rm esc}$, and defer exploring the impact of assuming mass and/or redshift dependent $f_{\rm esc}$ to future works.

\subsection{Thomson optical depth, \texorpdfstring{$\tau$}{tau}}
The optical depth is a measure of the scattering of CMB photons by free electrons produced by reionization. Given a reionization history, it is straightforward to obtain the Thomson scattering optical depth ($\tau$) as follows:
\begin{equation}
    \tau = \int_0^{\infty} \text{d}z \frac{c(1+z)^2}{H(z)}\, \sigma_T\, \bar{n}_{\rm H}\, [ x_{\rm HII}\,(1+\chi) +  \chi\,x_{\rm HeIII}],
\end{equation}
where $\sigma_T$ is the Thomson cross-section, and $c$ is the speed of light. 
The integrated optical depth is calculated considering hydrogen and helium reionization following \citet{Madau_2015}.
This optical depth can also be used to constrain the timing of reionization, where lower/higher $\tau$ corresponds to later/earlier reionization redshifts, respectively. 

In the right panel of Figure~\ref{fig:eor_con}, we show the resulting posterior distribution of fitting to the measured value of $\tau = 0.054 \pm 0.007$ from \citet{2020A&A...641A...6P}.
Similar to the case with $\dot N_{\rm ion}$ (left panel), $\tau$ alone does not place tight constraints on our parameters due to the large uncertainty that is consistent with a broad range of reionization histories. The tendency to favor models with high $f_{\rm esc}$ and negative $C$ is also seen. However, allowing much more negative $C$, as compared to values derived from $\dot N_{\rm ion}$ data suggests that a stronger contribution from low-mass halos is needed to reproduce the \citet{2020A&A...641A...6P} $\tau$ in our model. {The constraints placed on the $C$ parameter using $\rho_{\rm UV}$ data for $z\sim5$--$20$ {are}} consistent with constraints obtained by calibrating to \citet{2020A&A...641A...6P} $\tau$ at $1$--$2 \sigma$ level as shown by orange horizontal line.

\begin{deluxetable*}{cccccc}
\tabletypesize{\scriptsize}
\tablenum{3}
\tablecaption{Best-fit values of photon escape fraction ($f_{\rm esc}$) and power dependence on halo mass ($C$).}
\tablewidth{0pt}
\tablehead{
\colhead{Parameters} & \multicolumn3c{Constraints} & \colhead{Reionization (EoR)} & \colhead{Cosmic Dawn (CD)+Reionization (EoR)} \\
\colhead{} & \colhead{$\dot N_{\rm ion}(z=4.75)$} & \colhead{$x_{\rm HI}(z=5.9-7.6)$} & \colhead{$\tau$} & \colhead{$\dot N_{\rm ion}+x_{\rm HI}+\tau$} & \colhead{$\rho_{\rm UV}+\dot N_{\rm ion}+x_{\rm HI}+\tau$}
}
\startdata
\vspace{0.2cm} \\
${\bf f_{\rm esc}}$ & $0.68^{+0.23}_{-0.34}$  & $0.01^{+0.02}_{-0.01}$ & $0.68^{+0.22}_{-0.33}$ &  $0.02^{+0.02}_{-0.01}$ & $0.36^{+0.02}_{-0.02}$ \\ \\ \hline
\vspace{0.2cm} \\
${\bf C}$ & $-0.90^{+0.76}_{-2.13}$ & $0.73^{+0.22}_{-0.19}$ & $-2.31^{+1.43}_{-3.06}$ & $0.56^{+0.19}_{-0.17}$ & $-0.29^{+0.02}_{-0.02}$\\
\vspace{0.2cm}
\enddata
\tablecomments{The $A$ and $D$ parameters are fixed to values obtained by calibration to radiative transfer simulations ($A=10^{40}\,\Msun^{-1}\text{s}^{-1}$, $D=2.28$, see \citealt{2016MNRAS.457.1550H}), and $B$ to the derived value from fitting to $\rho_{\rm UV}$ in the entire redshift range $z\sim5$--$20$ ($\log (B/\Msun) = 7.67$, see Table~\ref{table:BandC}).}
\label{table:cdEoR}
\end{deluxetable*}

\begin{figure}
    \centering
    \includegraphics[width=0.45\textwidth]{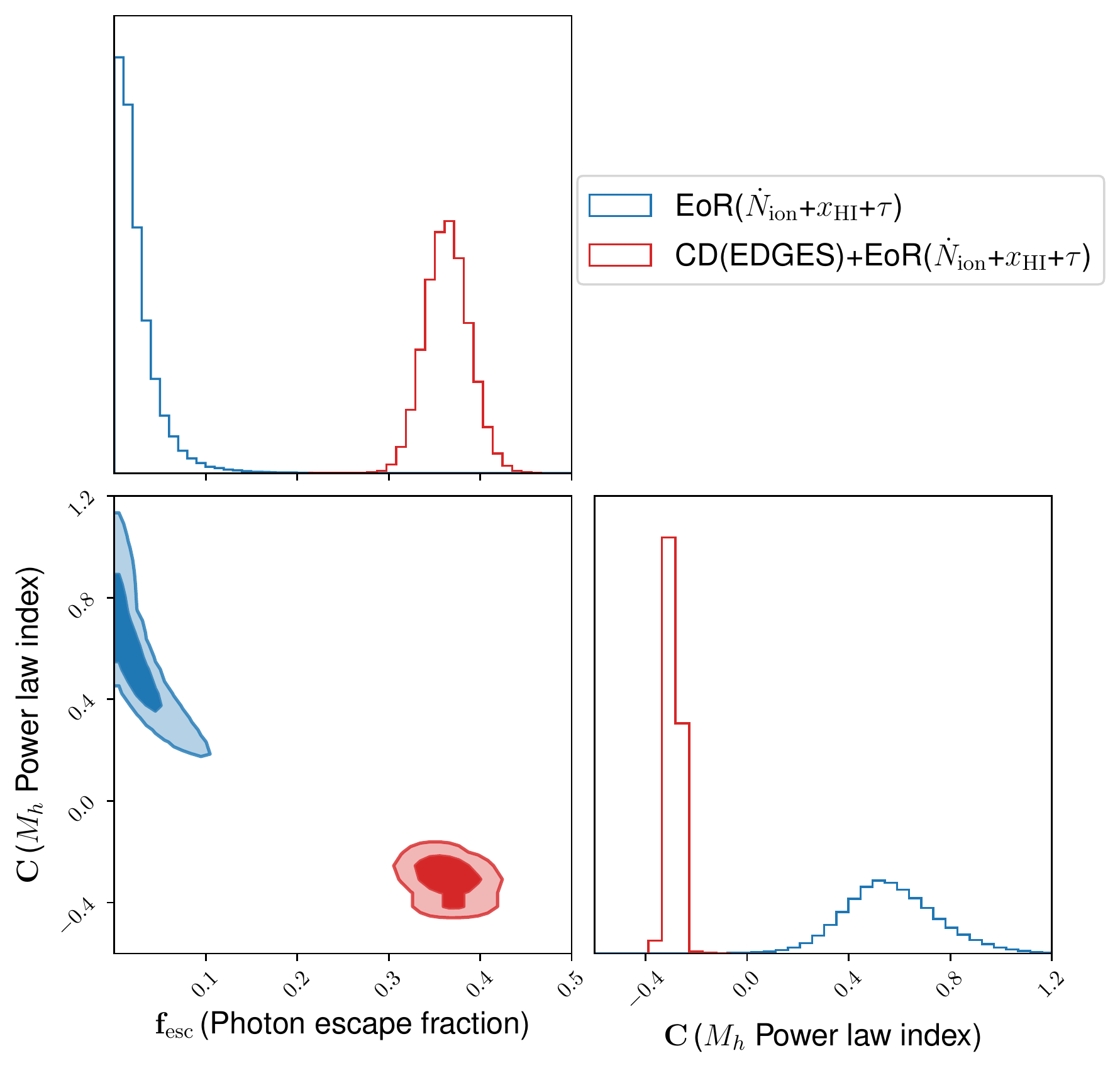}
    \caption{Comparison between the combined constraints from reioniztion, EoR ($\dot N_{\rm ion}, x_{\rm HI}$, $\tau$) in blue, and by adding cosmic dawn to reionization, CD ($\rho_{\rm UV}$ from EDGES) + EoR ($\dot N_{\rm ion}, x_{\rm HI}$, $\tau$) in red. The dark and light shaded contours correspond to the $1\sigma$ and $2\sigma$ levels, respectively. This demonstrates that adding cosmic dawn constraints shifts contours towards higher $f_{\rm esc}$ and negative $C$ parameters. This suggests that low-mass (faint) galaxies play a major role in bridging the gap between cosmic dawn and reionization.}
    \label{fig:cdEoR}
\end{figure}

\begin{figure}
    \centering
    \includegraphics[scale=0.65]{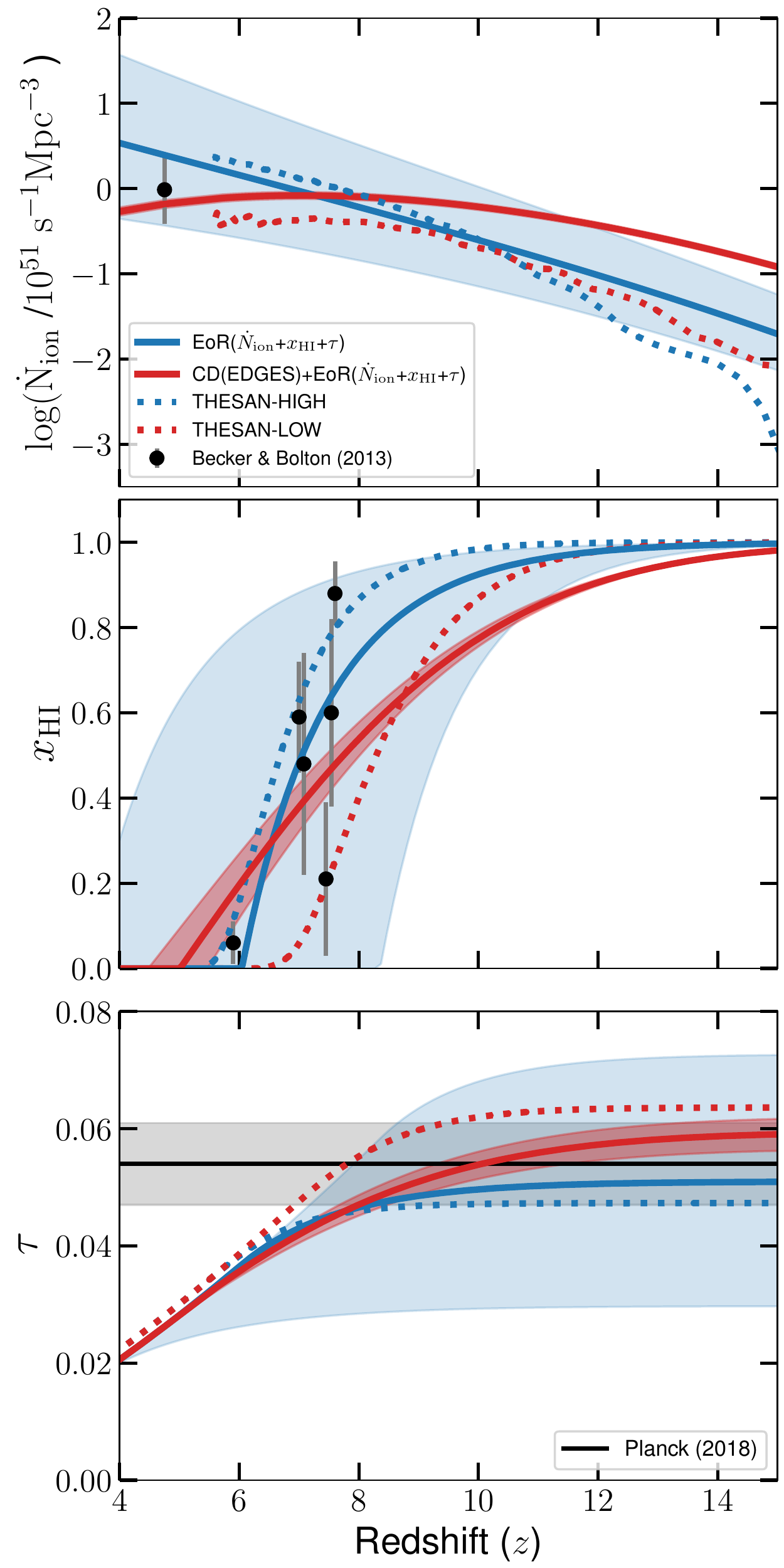}
    \caption{{\bf Upper:} Redshift evolution of the ionizing emissivity, $\dot N_{\rm ion}$, as predicted by our models, EoR (solid blue) and CD+EoR (solid red) along with the observational constraint by \citep{2013MNRAS.436.1023B}. Dotted blue and red curves show the \thesanhigh and \thesanlow simulations, respectively. {\bf Middle:} The neutral hydrogen fraction, $x_{\rm HI}$, as compared to
    data listed in Section~\ref{sec:xHI}.
    {\bf Lower:} The Thomson optical depth, $\tau$, as compared to the \citet{2020A&A...641A...6P} measurement (black line) with $1\sigma$ uncertainty (grey shaded region).
    Massive galaxies-dominated models (blue) favor a gradually increasing emissivity which in turn produces a later onset of reionization, more sudden and shorter duration, and lower optical depth. Faint galaxies-dominated models (red) favor a nearly flat emissivity evolution with the opposite characteristics.}
    \label{fig:pred}
\end{figure}

\section{Constraining the source model to reionization and cosmic dawn}
\label{sec:cdEoR}
We now turn our attention to the main goal of this work, which is to determine the range of models that can successfully reproduce reionization and cosmic dawn constraints. For reionization, we consider all measurements, from the previous section, which include the $\dot N_{\rm ion}$, $x_{\rm HI}$, and $\tau$ data. As mentioned earlier, the only existing constraint during cosmic dawn comes from the EDGES detection at $z \sim 17$, from which \citet{2018MNRAS.480L..43M} have been able to place constraints on the UV luminosity density $\rho_{\rm UV}$ (see \S\ref{sec:mcd}). We now constrain our source model to reionization only constraints ($\dot N_{\rm ion} + x_{\rm HI} + \tau$), and combined reionization and cosmic dawn constraints ($\dot N_{\rm ion} + x_{\rm HI} + \tau + \rho_{\rm UV}$). We refer to the former as EoR($\dot N_{\rm ion} + x_{\rm HI} + \tau$), and to the {latter} as CD(EDGES) + EoR($\dot N_{\rm ion} + x_{\rm HI} + \tau$). We compare the two resulting posterior distributions in Figure~\ref{fig:cdEoR}. Blue and red correspond to reionization only (EoR) and cosmic dawn plus reionization (CD+EoR), respectively. In both cases, combined observations place tight constraints on our parameters but in different parts of the parameter space. Comparing with Figure~\ref{fig:eor_con}, the reionization combined observations (blue, EoR) are mainly driven by the reionization history measurements ($x_{\rm HI}$). This shows that constraining to reionization observations favors models with low $f_{\rm esc}$ and positive $C$, indicating that massive galaxies play a major role in driving reionization. On the other hand, constraining to both reionization and cosmic dawn observations (red, CD+EoR) favors models with high $f_{\rm esc}$ and negative $C$, suggesting that low-mass galaxies play a dominant role to bridge the gap between these two epochs. This is represented by the diagonal shift from the blue to red contours once the cosmic dawn constraint ($\rho_{\rm UV}$ from EDGES) is added to the likelihood. All best-fit parameters and their 1$\sigma$ confidence intervals are listed in Table~\ref{table:cdEoR}.

In Figure~\ref{fig:pred}, we present predictions based on these best-fit parameters for the ionizing emissivity (top), reionization history (middle), and cumulative optical depth (bottom) from the EoR (blue) and CD+EoR (red) models. Shaded regions reflect the 1$\sigma$ uncertainty of the derived constraints (see Table~\ref{table:cdEoR}). In each panel, we add the relevant observations used in the MCMC analysis. We then compare our predictions with results from the \thesan project \citep{2022MNRAS.511.4005K, 2022MNRAS.512.4909G, 2022MNRAS.512.3243S}, which provides a suite of radiation-magneto-hydrodynamic simulations that self-consistently evolve reionization on large-scales ($L_{\text{box}} = 95.5\,\text{cMpc}$) and resolve the majority of ionizing sources responsible for it ($m_{\text{gas}} = 5.82 \times 10^5\Msun$) with galaxy formation physics based on the state-of-the-art IllustrisTNG model \citep{2014Natur.509..177V, 2014MNRAS.444.1518V, 2018MNRAS.480.5113M, 2018MNRAS.477.1206N, 2018MNRAS.475..624N, 2018MNRAS.475..676S, 2018MNRAS.475..648P}. Specifically, we compare with the \thesanhigh and \thesanlow simulation runs (dotted curves), where reionization is mainly driven by high-mass and low-mass galaxies, respectively.

In the upper panel of Figure~\ref{fig:pred}, we compare our predictions for the ionizing emissivity evolution with the \thesan runs \citep[based on the ray-tracing escape fraction calculations of][]{Yeh2022} and \citet{2013MNRAS.436.1023B} measurements. Our models (EoR, CD+EoR) and the different \thesan runs are consistent within the 1--2$\sigma$ levels of the observations. The emissivity evolution in the EoR model (blue) gradually increases towards decreasing redshifts as more massive halos form. This is a consequence of the positive $C$ value inferred by reionization observations, which leads to stronger contributions from massive halos. On the other hand, the CD+EoR model (red) infers a negative $C$ value, which gives more weight to the low-mass halos. This results in a higher $\dot{N}_{\rm ion}$ predicted by the CD+EoR model as compared to the EoR model predictions at high redshifts ($z \gtrsim 10$). As more massive halos form at lower redshifts ($z \lesssim 10$), the EoR model produces higher $\dot{N}_{\rm ion}$ than the CD+EoR model, where the emissivity in the latter starts to decrease due to the negative $C$ value. This leads to a flatter emissivity evolution in the CD+EoR model (red). Besides observations, our different models produce emissivity histories that are consistent with model predictions from the \thesan simulations. Our EoR and CD+EoR models show similar emissivity evolution to the \thesanhigh and \thesanlow, respectively, since the models share similar assumptions.

In the middle panel of Figure~\ref{fig:pred}, we show a reionization history comparison between our models and the \thesan runs against observationally inferred measurements. As with the emissivity, we see similar reionization histories between EoR/CD+EoR models and \thesanhigh/\thesanlow simulations, respectively. Reionization begins earlier in faint galaxies-dominated models (CD+EoR, \thesanlow) than in bright galaxies-dominated models (EoR, \thesanhigh), since {faint (bright) galaxies are predominant (rare) at high redshift}. Reionization is also more gradual (longer duration) in faint galaxies-dominated models, whereas bright galaxies-dominated scenarios yield a more sudden (shorter duration) neutral-to-ionized transition of the Universe.
This earlier onset of reionization in the CD+EoR and \thesanlow models translates to a higher cumulative optical depth, as seen in the bottom panel of Figure~\ref{fig:pred}, that is consistent within the 1--2$\sigma$ levels of the \citet{2020A&A...641A...6P} measurements. On the other hand, the sudden (late) reionization history in the massive galaxies-dominated models (EoR and \thesanhigh) produces a lower optical depth that is also consistent with the 1$\sigma$ level of the data. In addition,
our EoR model is in agreement with the late reionization models presented in \citet{2019MNRAS.485L..24K}, which reproduce the large-scale opacity fluctuations measurements \citep{2018ApJ...863...92B}.

\begin{figure}
    \centering
    \includegraphics[width=0.45\textwidth]{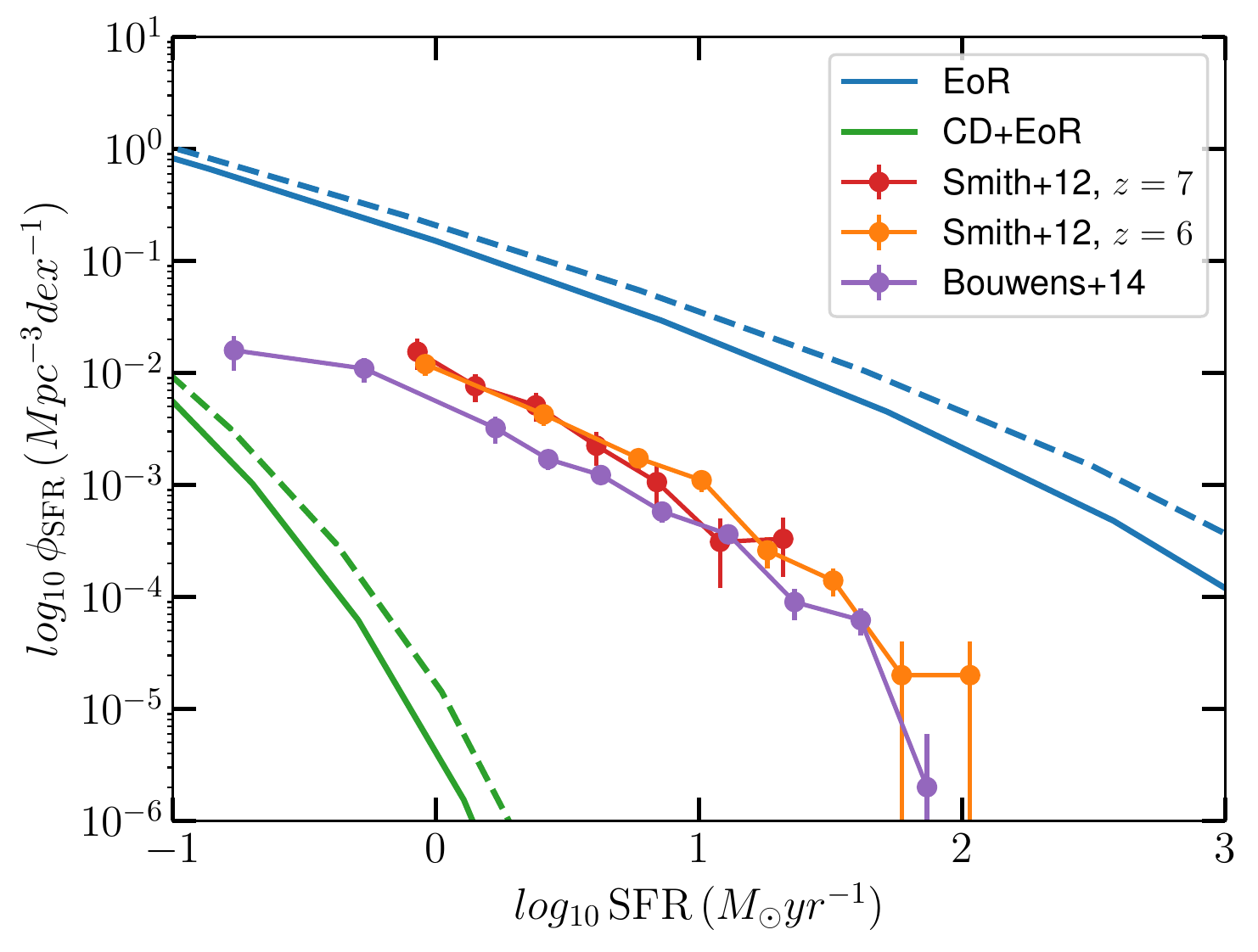}
    \caption{The star formation rate functions at $z=7$ obtained for EoR model and CD+EoR model are plotted by blue and green solid curves respectively. The dashed lines represent the same evaluated at $z=6$. The SFRFs from \citet{2012ApJ...756...14S} (at $z\sim6,7$) and \citet{2015ApJ...803...34B} (at $z\sim8$) are plotted, for comparison.}
    \label{fig:sfrf}
\end{figure}

To test whether our present models (EoR, EoR+CD) provide plausible predictions in terms of star formation, we calculate the star-formation efficiency, $f_{*}$ using Equations 1 and 4 in \citet{2016MNRAS.460..417S}. 
In case of EoR model, we find that it is $\sim 10\%$ at $M_{\text{h}} = 10^9\,\Msun$ and then increases with halo mass. In contrast, the efficiency peaks at $1\%$ for our CD+EoR model and then decreases with increasing halo mass as this model favors a greater 
contribution from faint galaxies. The $f_{*}$ evolution in \citet{2016MNRAS.460..417S} lies between the predicted $f_{*}$ in our models. In particular, at a representative redshift $z\sim8$ and at $M_{\text{h}} = 10^{10}\,\Msun$, our EoR model predicts efficiency of 40\% while the CD+EOR model predicts 0.5\%, that bracket the \citet{2016MNRAS.460..417S} model efficiency prediction of 3\%.
We further compute the star-formation timescales ($t_{\text{SF}}$) in our models using the approximate relation, $\rm SFR = f_* \frac{\Omega_b}{\Omega_m} \frac{M_h}{t_{\rm SF}}$. We find that it increases/decreases with $M_{\text{h}}$ in the EoR/EoR+CD models, respectively.
For both of our models, the average $t_{\text{SF}}$ is about $\sim 1\,\text{Gyr}$, consistent with commonly assumed values in traditional star formation models~\citep[e.g.][]{2003MNRAS.339..289S}.
However, it is worth mentioning that our models are in tension with instantaneous star formation rate functions (SFRF) or UV luminosity functions (UVLF) at different redshifts. The blue and green solid lines in Figure~\ref{fig:sfrf} represent the SFRF predictions from our EoR and CD+EoR models at $z=7$, while the dashed lines depict the same at $z=6$. For instance, we find that our EoR model overestimates the star formation rate functions over the entire SFR range as compared to \citet{2012ApJ...756...14S} (at $z\sim6$ and $7$) and \citet{2015ApJ...803...34B} (at $z\sim8$ ) SFRFs as shown in Figure~\ref{fig:sfrf}. In contrast, our CD+EoR model underestimates the SFRFs in efficiently star-forming and massive systems.
The SFRF data at $z=8$ in violet is taken from \citet{10.1093/mnras/stx2020}, which is calculated using the luminosity functions from \citet{2015ApJ...803...34B}.
As our models are calibrated to the globally averaged quantities, we do not expect our models to reproduce detailed redshift-instantaneous observables.
However, our CD+EoR model which is calibrated to all observables is broadly in good agreement with the \citet{2014ARA&A..52..415M} cosmic star formation rate densities at low redshifts.
As demonstrated by the CD+EoR model, it is entirely possible to reconcile the different observational constraints from the cosmic dawn to post-reionization using faint galaxies-dominated models of reionization, without the need to invoke new physics/exotic sources to explain the potential detection by EDGES.

\section{Conclusions}
\label{sec:conc}
We have presented a detailed analysis to explore the conditions by which the controversial EDGES detection \citep{2018Natur.555...67B} is consistent with reionization and post-reionization measurements including the ionizing emissivity ($\dot{N}_{\rm ion}$), IGM neutral hydrogen fraction ($x_{\rm HI}$), and Thomson optical depth ($\tau$) measurements. To account for the EDGES detection during cosmic dawn, we use the inferred constraint on the UV luminosity density ($\rho_{\rm UV}$) following \citet{2018MNRAS.480L..43M}, which is consistent with a simple extrapolation of deep Hubble Space Telescope (HST) observations of $4 < z < 9$ galaxies.

Using a semi-analytical framework of reionization with a physically-motivated source model \citep{2016MNRAS.457.1550H} coupled with MCMC likelihood sampling, our key findings are as follows:
\begin{itemize}
    \item Calibrating our source model $R_{\rm ion}$ to cosmic dawn constraints ($\rho_{\rm UV}$ from EDGES) favors models with negative $C$ values. This indicates that a stronger contribution from low-mass (faint) galaxies is required to reproduce the inferred $\rho_{\rm UV}$ constraints from EDGES (see Figure~\ref{fig:SFR} and Table~\ref{table:BandC}).
    
    \item Constraining our source model $R_{\rm ion}$ to different reionization observables is mainly driven by the IGM neutral fraction measurements where models with low $f_{\rm esc}$ and positive $C$ values are favored. This suggests that massive (bright) galaxies play the major role in driving reionization (see Figures~\ref{fig:eor_con} and \ref{fig:cdEoR} and Table~\ref{table:cdEoR}).
    
    \item Further constraining our source model $R_{\rm ion}$ to both reionization ($\dot{N}_{\rm ion}+x_{\rm HI}+\tau$) and the cosmic dawn constraint ($\rho_{\rm UV}$ from EDGES) favors models with high $f_{\rm esc}$ and negative $C$ values. This implies that low-mass (faint) galaxies would play a crucial role to bridge the gap between cosmic dawn and reionization (see Figure~\ref{fig:cdEoR} and Table~\ref{table:cdEoR}).
    
    \item Massive (bright) galaxies-dominated models produce an increasing emissivity that results in a later onset of reionization, more sudden and shorter reionization duration, and lower optical depth (see Figure~\ref{fig:pred}). Low-mass (faint) galaxies-dominated models result in a flatter emissivity evolution with the opposite reionization history characteristics.
\end{itemize}

It is worth mentioning several limitations to this work. First, the inferred $\rho_{\rm UV}$ values from EDGES compiled by~\citet{2018MNRAS.480L..43M} do not take into account the depth nor shape of the detected profile, but rather the constraints are imposed by the required Wouthuysen-Field coupling strength on UV radiation backgrounds at the detection redshift. As mentioned in~\citet{2018MNRAS.480L..43M}, accounting for the amplitude might require new physics, which in turn might alter our finding. We leave accounting for the whole properties of the detected profile to future works. Second, we use a linear relationship between SFR and $L_{\rm UV}$ in the entire redshift range ($z=5$--$20$), that has been derived from low redshift. This SFR--$L_{\rm UV}$ relation might not be linear at high redshift and assuming a different form might change our results quantitatively. Third, due to the high degeneracies between the $R_{\rm ion}$ parameters, we have fixed the amplitude $A$ and redshift dependence $D$ to values found in \citet{2016MNRAS.457.1550H} by calibrating to radiative transfer simulations of reionization. We have checked the results when $A$ is included as a free parameter and found relatively the same value obtained in \citet{2016MNRAS.457.1550H}. The parameter $D$ might not be the same at different redshifts due to evolution in galactic feedback. However, we have obtained our fit from radiative transfer simulations using data from $z=6$--$12$, and it has been shown that the observed SFR function parameters show a weak dependence on redshift in \citet{2012ApJ...756...14S}. Hence, we do not expect a qualitative change in our results if $D$ is varied. Fourth, the quantitative results might be different if the $A$, $B$, and $D$ parameters were fixed to different values. Nevertheless, the qualitative result (high/low redshift data prefers models with more negative/positive $C$ and higher/lower $f_{\rm esc}$, respectively) would be similar. {Fifth, our models have been adjusted to match the globally averaged quantities. {As a result, we observe that the EoR and CD+EoR model over-produces and under-produces the star formation rate functions (or equivalently UV luminosity functions) respectively.}}  We leave to future works to perform a detailed analysis to calibrate our models to all global and redshift-instantaneous quantities, including the SFRF and UVLF at different redshifts using the recently high redshift data by JWST and the previous HST low redshift data.

It is worth noting that calibrating our models to a different $\tau$ value, $\tau = 0.0627^{+0.0050}_{-0.0058}$, as recently determined by \citet{2021MNRAS.507.1072D}, and to the updated measurements of $\dot N_{\rm ion}$, presented by \citet{2021MNRAS.508.1853B}, does not alter our findings.
In summary, our results demonstrate that it is entirely possible to reproduce both cosmic dawn and reionization constraints with faint galaxies-dominated models without requiring new physics or exotic sources as our models with deduced parameters align well with the star formation efficiencies and time scales that have been reported in previous literature. Our results shed additional light on the roles of faint and bright galaxies during cosmic dawn and reionization, which can be tested by upcoming JWST surveys.

\section{Acknowledgements}
The authors acknowledge insightful discussions with Adam Lidz, Tirthankar Roy Choudhury, Guochao Sun, Greg Bryan, Romain Teyssier, Lars Hernquist, Hy Trac, Matt Orr, Ulrich Steinwandel, Shy Genel, Francisco Villaescusa-Navarro, Kanan Datta, Saumyadip Samui and Ken Van Tilburg which have improved the paper significantly. We thank the anonymous referee for the comments which have improved the paper quality greatly. AB acknowledges financial support from University Grant Commission (UGC), Govt. of India as a senior research fellow. A large fraction of this work has been done at CCA, Flatiron Institute, and AB, SH acknowledge support provided by the Simons Foundation. SH also acknowledges support for Program number HST-HF2-51507 provided by NASA through a grant from the Space Telescope Science Institute, which is operated by the Association of Universities for Research in Astronomy, incorporated, under NASA contract NAS5-26555. This work also used the Extreme Science and Engineering Discovery Environment (XSEDE), which is supported by National Science Foundation grant number ACI-1548562, and computational resources (Bridges) provided through the allocation AST190009.

\bibliography{sample631}{}
\bibliographystyle{aasjournal}



\end{document}